\documentclass[a4paper,11pt]{article}
\pdfoutput=1 

\usepackage{jheppub} 
\usepackage[T1]{fontenc} 

\usepackage{blindtext}
\usepackage[skins,theorems]{tcolorbox}
\usepackage{empheq}
\usepackage[export]{adjustbox}
\usepackage{caption}

\usepackage{booktabs} 
\usepackage[utf8]{inputenc}
\usepackage[english]{babel}
\usepackage{amsfonts}
\usepackage{ragged2e}
\usepackage{etoolbox}
\usepackage{lipsum}
\usepackage{multirow}
\usepackage{tcolorbox}
\usepackage{latexsym}
\usepackage{enumitem}
\usepackage{comment}
\usepackage{mathtools}

\newcounter{saveenumi}

\newcommand{\del}{\partial}
\newcommand{\Lie}[1]{\mathcal{L}_{#1}\,}
\newcommand{\wn}{\widetilde\nabla}
\newcommand{\ep}{\epsilon^{\mu\nu\rho}}

\title{\boldmath A (1+1)-dimensional Lifshitz Weyl Anomaly From a Schr$\mathrm{\ddot{o}}$dinger-invariant Non-relativistic Chern-Simons Action}
\author{Amr Ahmadain}
\affiliation{Department of Physics, University of Virginia, \\ Charlottesville, VA 22904, USA}
\emailAdd{aaa9aj@virginia.edu}

\abstract{The main result of this paper is that the Weyl anomaly of a $z=2$ (1+1)-dimensional Lifshitz effective action can be derived from a (2+1)-dimensional non-relativistic Schr$\mathrm{\ddot{o}}$dinger-invariant Chern-Simons (NRSCS) action which was shown to be equivalent to a specific Weyl-invariant non-projectable Horava-Lifshitz action of gravity. On a manifold with a boundary, we will show that the (1+1)-dimensional Lifshitz Weyl anomaly can be derived from a specific term, the torsional CS (tCS) term, in the NRSCS action built from the gauge fields of the Weyl and special conformal symmetry generators of the centrally-extended Schr$\mathrm{\ddot{o}}$dinger algebra. We also focus on the $z=1$ Lifshitz Weyl anomaly and attempt to elicit its geometric and physical nature, in particular its relationship with the Lorentz anomaly of a (1+1)-dimensional CFT effective actions. We show that it is directly related to the curvature scalar of the dual Lorentz connection $d\star\omega$, the integral of which is known to be a topological invariant.  We also point out that making the anomalous Lifshitz quantum effective action Weyl-invariant amounts to obtaining the equation of motion for a stationary chiral boson which happens to be the spatial-component of the acceleration vector. By putting boundary conditions on the spatial slices, the time dependence of the lapse function in the Arnowitt, Deser and Misner (ADM) decomposition is eliminated and the result is a Rindler metric. We finally discuss several issues related to the (1+1)-dimensional Lifshitz Weyl anomaly regarding edge physics of fractional quantum Hall states and anomaly cancellation by anomaly inflow. 
}

\begin{document} 
	\maketitle
	\flushbottom

	\section{Introduction}
	Anomalies are symmetries of the classical action that are broken at the quantum level. Gravitational anomalies of one-loop quantum effective actions arise after coupling classical field theories to curved background geometry and integrating out all dynamical fields in the partition function. Since the quantum energy-momentum tensor, by definition, encodes the response of effective actions to infinitesimal variations in the underlying background metric, they are the central objects in studying gravitational anomalies. In particular, a gravitational conformal (or Weyl anomaly) is the statement that the quantum effective action is not invariant under local rescaling of the background metric. The trace of the expectation value of the energy-momentum tensor is the canonical test of whether the theory is Weyl anomalous or not. If the trace is non-zero, the quantum theory suffers a conformal anomaly.
	
	Recently, there has been a considerable level of activity in studying Weyl anomalies in non-relativistic field theories \cite{Gomes2012, Boer2012,Oz2014,Nardelli2016, Oz2016, Filip2016,Moshe2017,Mitra2017,Grinstein2017,Nardelli2017}. Non-relativistic field theories do not place space and time on an equal footing and thus introduce a degree of anisotropy between them. Lifshitz field theories are important examples of non-relativistic theories which are locally symmetric under foliation-preserving diffeomorphism (FPD), as opposed to full diffeomorphism invariance in relativistic theories of gravity, and \textit{anisotropic} Weyl scaling transformations characterized by a dynamical scaling exponent $z$. In FPD-invariant theories, the spacetime is naturally foliated into \textit{equal-time slices} with a smooth timelike 1-form $n_{\mu}$ \textit{normal} to the foliation leaves which is normalized by the spacetime metric $g_{\mu\nu}$. Along with Schrodinger field theories, they are used to study and characterize several condensed matter systems near or at the quantum critical points \cite{Henkel94,Fradkin04,Troyer11}. 
	
	Studying quantum anomalies of non-relativistic field theories typically requires coupling to non-relativistic geometries. Newton-Cartan (NC) spacetime with a \textit{torsion} tensor, or torsional NC geometry (TNC) has recently been the focus of intense study. TNC geometry has appeared in different physical setups, for example, in boundary effective actions of non-relativistic holographic theories \cite{Rollier14,Rollier14-2,Obers2015} and in effective field theories of quantum Hall states \cite{Son13,Wu2015}. Weyl-invariant field theories coupled to flat NC spacetime were constructed in \cite{Obers14,Obers15}. 
	
	Recently, Weyl anomalies of Lifshitz field theories coupled to NC geometry with \textit{temporal torsion}, where the 1-form $n_{\mu}$ satisfies the Frobenius condition, $n\wedge dn = 0$, have been calculated in several spacetime dimensions and for multiple values of the scaling exponent $z$ by solving the Wess-Zumino consistency condition \cite{Oz2014}. It was found in \cite{Oz2014} that while the conformal anomalies of (1+1)-dimensional relativistic conformal field theories are \textit{type-A}, those of Lifshitz field theories belong to \textit{type-B}. Anomalies in the latter class satisfy \textit{trivial descent} equations, or equivalently, consist of Weyl-invariant scalar densities with effective actions that are scale-dependent.
	
	In 1+1 dimensions, which is the focus of this paper, and for \textit{any} value of $z$, only one \textit{trivial descent} anomaly i.e. a trivial descent cocyle modulo a coboundary term, was found in the parity-odd, mixed-derivative sector of the 1+1 Lifshitz cohomology of the relative Weyl operator \cite{Oz2014}. The rest of the cocycles were shown to be trivial descent \textit{coboundaries} and thus, can be removed by local counterterms. Also, the (1+1)-dimensional Weyl anomaly breaks time-reversal invariance. Using the ADM coordinates, we will illustrate that the 1+1 Lifshitz Weyl anomaly can be directly interpreted as the curvature of the torsion 1-form $ a_\mu$ in the TNC geometry. In terms of the lapse function $N(x,t)$ of the ADM parametrization, we will see that the anomalous degree of freedom is a direct consequence of the \textit{time-dependence} of the lapse function. Hence, the presence of this 1+1 Lifshitz Weyl anomaly necessarily implies that energy is not conserved in the system. More concretely, we will show that the $x$-component of the torsion (or acceleration) vector of the NC geometry defined by $a_x(x,t) = \frac{\del_xN(x,t)}{N(x,t)}$, is not conserved as a result of the Weyl anomaly, and hence it physically represents \textit{jerk} in the system. 
	
	In a separate yet related track, Horava-Lifshitz (HL) theories of gravity have been introduced as a power-counting renormalizable non-relativistic quantum theory of gravity with anisotropic scaling symmetry \cite{Horava1-09,Horava2-09}. The key idea behind HL gravity theories is that by introducing terms with higher spatial derivatives, the ultraviolet (UV) behavior of the graviton propagator is improved and the theory eventually becomes power-counting renormalizable. When the number of spatial dimensions equals the dynamical scaling exponent $z$, Weyl-invariant actions can be found. HL actions break the principle of general covariance by foliating spacetime with space-like surfaces and introducing extra geometric data that affect the number and dynamics of degrees of freedom in the theory. As a result, not only do they describe the dynamics of the helicity-2 modes of the spatial metric but also an extra \textit{helicity-0} scalar mode. Since this foliation mode is an excitation of the global time, it is usually called a \textit{scalar khronon} \cite{Blas2011}. Therefore, it is natural to expect that gravitational Weyl anomalies of Lifshitz quantum effective actions coupled to background NC geometry with temporal torsion will somehow encode this extra foliation structure. In fact, this is precisely what 1+1 Lifshitz Weyl anomaly encodes: the time derivative of the lapse function in the ADM (or unitary gauge) or two time derivatives of the khronon field in a general coordinate system \cite{Blas2011}. It is well known that the breaking of general covariance in HL gravity theories leads to infrared instablities which has cosmological implications as shown in \cite{Sunny16}.
	
	The connection between dynamical NC geometry, with and without torsion, to HL gravity theories was demonstrated in \cite{ObersDynmicTNC2015}. More specifically, it was shown that dynamical NC geometries without torsion gives rise to projectable HL gravity while those with twistless torsion (TTNC) i.e. those that obey the Frobenius condition and do not allow torsion on the spatial slices, give rise to the non-projectable version of HL gravity. Projectable HL gravity theories are those where the lapse function in the ADM decomposition of spacetime is only dependent on time, i.e. $ N(t) $ whereas the non-projectable version emerges when it is a function of both space and time, i.e. $ N(x,t) $ and hence contain the acceleration vector as a dynamical quantity. Weyl-invariant theories of HL gravity can only be non-projectable \cite{Thompson2012}. Gauging a symmetry algebra is tightly related to spacetime geometry. Just as gauging the Poincare algebra gives rise to Riemannian geometry that couples to relativistic field theories, it was shown in \cite{Roo2011} and \cite{Rosseel2015} that gauging the Bargmann and Schrodinger algebras leads to NC geometries without and with torsion respectively. More specifically, as noted in \cite{ObersDynmicTNC2015}, adding torsion to the NC geometry amounts to making it locally scale-invariant by gauging the Schrodinger algebra. Therefore, it stands to reason that the 1+1 Lifshitz anomaly is directly linked to the torsion vector of the NC geometry, which as shown in \cite{ObersDynmicTNC2015}, maps directly to the acceleration vector in HL gravity theories.
	
	
	By gauging the non-relativistic Bargmann and centrally-extended Schrodinger algebras, the authors in \cite{ObersNRSCS} constructed a (2+1)-dimensional non-relativistic Bargmann-invariant and Schr$\mathrm{\ddot{o}}$dinger-invariant Chern-Simons (NRSCS) actions, respectively. While the former gives projectable HL theory of gravity, the latter, which is the focus of this paper, is equivalent to $z=2$ conformal i.e Weyl-invariant non-projectable HL gravity. CS actions are known to be gauge-invariant up to total derivative terms. On manifolds with boundaries, these total derivative terms can generate anomalies of boundary quantum effective actions. For example, in the context of AdS/CFT, under a diffeomorphism or Lorentz transformation, the boundary term of the gravitational CS (gCS) action added to a three-dimensional on-shell gravitational action generates a diffeomorphism or  Lorentz anomaly, respectively, of a two-dimensional \textit{boundary} CFT effective action \cite{Larsen2006}. Analogously, by placing the NRSCS action on a manifold with a boundary, it will be shown in this paper that under a Weyl transformation, the NRSCS action changes by a total derivative term that precisely matches the boundary Weyl anomaly of a $z=2$ Lifshitz effective action coupled to background TTNC geometry. This is the main result of this paper. More concretely, we will show that the 1+1 Lifshitz Weyl anomaly can be derived holographically from a specific term in the three-dimensional NRSCS action constructed from the gauge fields of the Weyl and special conformal symmetry generators of the Schrodinger algebra. Throughout this paper, we call this term the torsional CS (tCS) term. We will show that the tCS term added to a three-dimensional Weyl-invariant HL gravity action plays a role similar to what the gCS term plays when added to a three-dimensional diffeomorphism-invariant action.
	
	We also focus in this paper on the $z=1$ Lifshitz Weyl anomaly, where space and time scale \textit{relativistically}. This anomaly possesses some interesting properties. In addition to being universal i.e. a $z$-independent anomaly in 1+1 dimensions, it was shown that it is the \textit{Weyl partner} of the Lorentz anomaly of 1+1 CFT effective actions \cite{Oz2014}. To explicitly demonstrate the latter property, the authors shifted the Lorentz anomaly of a 1+1 CFT effective action $W_{\text{conf}}[e^a_{\mu}]$ where $e^a_{\mu}$ are the local orthonormal frame fields, into foliation dependence, i.e. to \textit{dependence} on $n_\mu$ of the background foliation geometry. By doing so, the original CFT effective action $W_{\text{conf}}[e^a_{\mu}]$ turns into a $z=1$ FPD-invariant Lifshitz theory and thus the physical content of the anomalous Ward identity, the expectation value of the trace of the energy-momentum tensor with respect to $e^a_{\mu}$,  $\left\langle T^\mu_{(e)\mu}\right\rangle$, now reflects a Lifshitz Weyl anomaly rather than a Lorentz anomaly of the original CFT effective action. In other words, the anomalous local frame rotations of $W_{\text{conf}}[e^a_{\mu}]$ is exchanged for anomalous Weyl transformations of the foliation geometry.
	
	One of the goals of this paper is to further elicit the unique relationship between the $z=1$ Lifshitz Weyl anomaly and Lorentz anomaly in 1+1 dimensions. First, we will illustrate that the $\left\langle T^\mu_{(e)\mu}\right\rangle$ can be interpreted as the scalar curvature of the \textit{dual} Lorentz connection 1-form in a local two-dimensional gravity theory with conformal and Lorentz anomalies constructed in \cite{Solod90}. In addition, we will show that restoring the local Weyl symmetry of the Weyl-anomalous effective action $W[e^a_{\mu}]$, where $e^a_{\mu}$ are the frame fields, eliminates the gauge redundancy associated with the time dependence of the lapse function and thus makes $a_x$ a conserved quantity. Insisting on the Weyl invariance of $W[e^a_{\mu}]$ amounts to solving an equation of motion for a chiral boson at rest after imposing appropriate appropriate boundary conditions at the spatial boundaries. We show that solving this equation yields the Rindler metric for hyperbolically accelerated observers. 
	
	This paper is organized as follows. In Section \ref{NC_Geom}, we give the basic definitions of NC geometry with temporal torsion required for the discussion in the paper. In Section \ref{ADMParam}, after introducing the ADM coordinates, we define the torsion and curvature tensors and give the expression of the anomalous Ward identity for the Lifshitz Weyl symmetry. In Section \ref{LifWeylAnomGeom}, we use the ADM gauge to elicit the geometric properties of the 1+1 Lifshitz Weyl anomaly. In Section \ref{AnomalyDerivation}, we proceed to demonstrate the main result of this paper where we will show how the 1+1-dimensional Lifshitz Weyl anomaly can be derived from the torsional Chern-Simons (tCS) term in (2+1)-dimensional NRSCS action and how the tCS term added to the Weyl-invariant HL gravity action plays a role similar to what the gCS term plays when added to a three-dimensional diffeomorphism-invariant action. In Section \ref{Weyl-Lorentz-Anom-Relation}, we shift our focus on the $z=1$ Lifshitz Weyl anomaly where we will first review its connection to the Lorentz anomaly in 1+1 CFT effective action as pointed out in \cite{Oz2014} and then show how it is related to the scalar curvature of the dual Lorentz connection in a local two-dimensional gravity theory with conformal and Lorentz anomalies. In Section \ref{AnomalyCancellation}, we discuss how canceling the Weyl anomaly leads to an equation of motion for a stationary chiral boson which by solving, we obtain the Rindler metric. We also use Darboux's coordinates to explain how as a result of canceling the Weyl anomaly, we get a symplectic manifold with a \textit{Hamiltonian} function and Hamiltonian vector field. In Section \ref{Discussion}, we discuss several directions for future work and open questions.

	\section{The (1+1)-dimensional Lifshitz Weyl Anomaly}\label{Section2}
	In this section, after very briefly reviewing the basic structure of the NC spacetime geometry with temporal torsion in Section \ref{NC_Geom}, we introduce the ADM parametrization in Section \ref{ADMParam} and then define the anomalous Lifshitz Ward identity before we get to discussing the geometric nature of the Weyl anomaly, true for $z\geq 1$, in Section \ref{LifWeylAnomGeom}. In Sections (\ref{Weyl-Lorentz-Anom-Relation}) and (\ref{AnomalyCancellation}), we focus on the $z=1$ case. Throughout this section, the indices $\mu, \nu$ refer to spacetime coordinates and $i,j$ to spatial ones. For the local tangent frame coordinates i.e. vilbeins, $a, b, \dots = 0,1, \dots d$ and $A, B = 1, \dots d$ are used to denote spacetime and spatial indices, respectively. We closely follow the notations and conventions used in \cite{Oz2014}.
	
	\subsection{The Newton-Cartan (NC) Geometry} \label{NC_Geom}
	NC geometry, as opposed to Riemannian geometry in relativstic theories, is what couples naturally to non-relativistic field theories. Anomalies in non-relativistic quantum field theories coupled to NC geometry are prime examples of how the geometrical objects of the NC geometry become manifest. In non-relativistic field theories, the \textit{time} direction plays a major role and spacetime is naturally foliated into \textit{equal-time slices} or surfaces of simultaneity. Assuming the existence of a scalar field globally defined on the foliated spacetime manifold, a smooth local vector field $t_{\mu}$ \textit{normal} to the foliation leaves and normalized by the spacetime metric $g_{\mu\nu}$ given by $ n_\mu = t_\mu /\sqrt{|g^{\beta\gamma} t_\beta t_\gamma|} $ is the most basic and intrinsic geometrical object defined on this manifold. Tangent vector fields to the foliation leaves are then defined as the kernel of $ n_\mu $ i.e. they satisfy $n_\mu V^\mu = 0 $.
	
	More formally, the basic geometrical structure on a $D=$ (\textit{d}+1)-dimensional NC manifold $M$ consists of an everywhere smooth \textit{temporal} metric $t_\mu t_\nu$, a \textit{degenerate} symmetric spatial component $h^{\mu\nu}$ with signature $(0,+, \dots, +)$ i.e. corank-1 tensor \cite{Geracie15} and a notion of a covariant derivative $\nabla$ all satisfying the following constraints
	\begin{equation}
	h^{\mu\nu}t_{\mu} = 0 , \quad  t_{\mu}t^{\nu} = -1, \quad \nabla_{\mu}t_{\nu}  = \nabla_{\mu}\gamma^{\nu \lambda} = 0  \,.
	\end{equation}
	While the 1-form $n_{\mu}$ provides a notion of a \textit{clock}, its inverse $n^{\mu}$ denotes the direction of time often called the velocity field. Using the NC geometrical objects defined above, one can construct a non-degenerate symmetric rank-2 tensor $g_{\mu \nu}$ with a Lorentzian signature (-1,1,\dots 1) that has a temporal component $n_\mu n_\nu$ as well as a spatial component $\gamma_{\mu \nu}$, i.e. $g_{\mu \nu} = \gamma_{\mu \nu} - n_\mu n_\nu$. Note that $g_{\mu \nu}$ is not a Lorentzian metric as it would normally be in a relativistic theory. For a more formal and thorough definition of the NC spacetime, see \cite{Karch14,Geracie15}.
	
	Following \cite{ObersDynmicTNC2015}, there are three different constraints on the foliation 1-form $ n_{\mu} $ that each give a different type of NC geometry:
	\begin{enumerate}[label=(\roman*)]
		\item \textit{Torsionless} NC geometry: \textit{$ dn = 0 $} where the connection $\Gamma^{\lambda}_{\mu \nu} $ is torsionless 
		\item  \textit{Twistless Torsion} Newton-Cartan (TTNC) or \textit{temporal torsion} geometry
		\begin{equation} \label{TTNCCondition}
		\partial_{\mu}n_{\nu} - \partial_{\nu}n_{\mu} = a_{\mu}n_{\nu} - a_{\nu}n_{\mu}\, ,
		\end{equation}
		where the \textit{acceleration or torsion vector} $a_{\mu}$ is defined as the Lie derivative of the foliation 1-form along the $ n^{\mu} $ velocity vector field  $ n_{\mu} $
		\begin{equation}
		a_{\mu} = \mathcal{L}_{n}n_{\mu} \, ,
		\end{equation}
		
		The TTNC constraint in (\ref{TTNCCondition}) is an expression of the solution of the \textit{Frobenius condition}, an integrability condition that states a \textit{local} 1-form defines a codimension-1 foliation if and only if it satisfies the following constraint:
		\begin{equation} \label{FrobCon}
		n \wedge dn = 0\,.
		\end{equation}
		Imposing the Frobenius condition means that it is always possible to find a coordinate system in which the spacetime manifold is foliated into equal-time hypersurfaces or foliation leavs $\Sigma_t$ to which the unit time-like vector field  $n_{\mu}$ is orthogonal. The Frobenius condition makes the TTNC spacetime \textit{causal} in the sense that if it does not hold, then each point $p \in M$, has a neighborhood within which all points are spacelike separated. It is also important to mention that TTNC spacetimes, while being causal, still lack the notion of an \textit{absolute} time measured by all observes along their worldlines. The difference between the total coordinate time measured by two observers starting at different points on $\Sigma_{t_1}$ and traveling to another time slice $\Sigma_{t_2}$ along their respective wordlines is exactly the torsion in \textit{time} $dn = a\wedge n$ \cite{Geracie15}. This point is key to understanding the physical as well the geometrical meaning of the Weyl anomaly.
		
		\item \textit{Torsional} NC or TNC geometry where $ n_{\mu} $ is not constrained and has therefore arbitrary torsion.
	\end{enumerate}
	We will later see how the (1+1)-dimensional Lifshitz Weyl anomaly is directly related to the TTNC geometry.

	\subsection{The ADM Parametrization} \label{ADMParam}
	In the ADM decomposition, one chooses coordinates $(t,x^i)$ such that the leaves of the foliation are given by constant-time slices $t=const$ and $ x^i $ for the coordinates in each leaf. The ADM metric assumes a frame, the \textit{unitary or synchronous} gauge where the time of the \textit{spatial} foliation hypersurfaces coincides with coordinate time $t$ such that $n_t = N(x,t)$ and $ n_i = 0 $ and the result is a spacetime metric with a well-defined notion of global time.  In this gauge, the ADM metric describes the TTNC geometry where the Frobenius condition given in (\ref{FrobCon}) is \textit{automatically} satisfied. In these \textit{preferred} coordinates, the metric $g_{\mu \nu} = \gamma_{\mu \nu} - n_\mu n_\nu$ takes the form \footnote{For details on how the ADM metric is obtained from the fundamental NC objects in the process of gauging the Bargmann algebra, see section 8 in \cite{ObersDynmicTNC2015}. }
	\begin{equation}\label{MetricADM}
	g_{\mu\nu}= 
	\begin{pmatrix}
	g_{tt}	&  g_{ti} \\ 
	g_{jt}	& g_{ii}
	\end{pmatrix} 
	=
	\begin{pmatrix}
	-N^2 + N^i N_i	&   N_i \\ 
	N_j &  h_{ij}
	\end{pmatrix} \,,
	\end{equation}
	while the components of the inverse metric are given by
	\begin{equation}\label{MetricADM}
	g^{\mu\nu}= 
	\begin{pmatrix}
	g^{tt}	&  g^{ti} \\ 
	g^{jt}	& g^{ii}
	\end{pmatrix} 
	=
	\begin{pmatrix}
	-\frac{1}{N^2}	&  \frac{N^i}{N^2} \\ 
	\frac{N^j}{N^2} &  h^{ij} - \frac{N^iN^j}{N^2}
	\end{pmatrix} \, ,
	\end{equation}
	where $h_{ij} $ is the induced metric on the foliation leaves, $ N^i $ is the shift vector and $ N(x,t) $ is the lapse function.
	The covariant volume element in these coordinates is given by:
	\begin{equation}
	\sqrt{-g}\,d^{(d+1)}x = N\sqrt{h}\,dt\,d^d x .
	\end{equation}
	The timelike normal to the foliation is given by
	\begin{equation}\label{ADMTimelikeFol}
	\begin{split}
	& n_\mu = N(-1, 0) \, ,\\
	& n^\mu = \frac{1}{N} (1,-N^i)\, .
	\end{split}
	\end{equation}
	In $d$ spatial dimensions, the Lie derivative of a foliation tangent tensor $V_{ijk\dots} $ is given by
	\begin{equation}\label{LieADM}
	\mathcal{L}_n V_{ijk\dots} = \frac{1}{N} \partial_t V_{ijk\dots} -\frac{1}{N} \mathcal{L}^{(d)}_{\vec N} V_{ijk\dots} \,,
	\end{equation}
	where $ \mathcal{L}^{(d)}_{\vec N} $ is the Lie derivative inside the foliation leaf taken along the direction of the shift vector $ N^i $.
	\\~\\
	\noindent In 1+1 dimensions, the extrinsic curvature tensor is simply given by
	\begin{equation}\label{ExtrCurvADM}
	K_{xx} = \frac{1}{2N} (\partial_t h_{xx}) \,.
	\end{equation}
	whereas the $x$-component of the acceleration vector in 1+1 dimensions is given by
	\begin{equation}\label{TorsionADM}
	a_x = \frac{\del_x N}{N}\,.
	\end{equation}
	and the  temporal component is $a_t= N^x a_x$.
	In 1+1 dimensions, with a non-zero shift vector $N^x$, the ADM metric is given by \footnote{By working in the ADM preferred coordinates, the shift vector $N^x$ can always be removed by an FPD transformation.}
	\begin{equation}\label{ADMMetric}
	ds^2 = -N^2\,dt^2 + N_x\, dxdt + h_{xx} \, dx^2 \,.
	\end{equation}
	We will later discuss the consequences of having a non-zero $a_t$ with a constant $N^x$. In 1+1 spacetime dimensions with a zero shift vector, the spacetime metric has only two degrees of freedom: the lapse function $ N(x,t) $ and the spatial metric $h_{ij}$. The spatial metric $ h_{ij} $ is a rank-0 tensor, i.e. a function $ h_{xx}(x,t) $. The volume element $dV$ is then $ \sqrt{-g}\,dt\,dx = N\sqrt{h}\,dt\,dx $ .

	\subsection{The 1+1 Weyl Anomaly And Anomalous Ward Identity}\label{LifWeylAnomGeom}
	In this section, we attempt to illustrate the geometric nature of the 1+1-dimensional Lifshitz Weyl anomaly and how it is closely related to the NC geometry with temporal torsion. To that effect, we use the ADM coordinates to define some basic TTNC objects required to understand the geometric nature of the 1+1 Weyl anomaly. As mentioned in the introduction, dynamical TTNC gives rises to non-projectable Horava-Lifshitz theory of gravity. Since this approach is useful for our purposes in this section, we use some of the definitions in \cite{Obers2015} and \cite{ObersDynmicTNC2015}.
	
	The antisymmetric part of the torsion tensor $\Gamma^{\lambda}_{\mu \nu}$, is expressed as
	\begin{equation}\label{TorsionSpinConn}
	\Gamma^{\lambda}_{[\mu \nu]} = n^{\lambda}\partial_{[\mu}n_{\nu]} =  n^{\lambda}a_{[\mu}n_{\nu]} =  n^{\lambda}R_{\mu\nu}(H)\, ,
	\end{equation}
	where $R_{\mu\nu}(H)$ is the curvature 2-form of $n_{\nu}$ defined as the \textit{gauge field} of the generator of time translation symmetry, i.e. the Hamiltonian $(H)$
	\begin{equation} \label{CurvatureFoliation}
	R(H) = (\partial_{\mu}n_{\nu} - \partial_{\nu}n_{\mu}) \, dx^{\mu}\wedge dx^{\nu} \,.
	\end{equation}
	In 1+1 spacetime dimensions, imposing the Frobenius condition and using the ADM gauge, the only non-vanishing component of the torsion 2-form $\Gamma$ as defined by equation (2.27) in \cite{ObersDynmicTNC2015} is given by
	\begin{equation} \label{TraceTorsionTensor}
	\Gamma = \frac{1}{2}\Gamma^{t}_{[x t]} \,dx\wedge dt =  n^{t}(a_{x}n_{t} - a_{t}n_{x}) \,dx\wedge dt = a_{x} \,dx\wedge dt = \frac{\partial_x N(x,t)}{N(x,t)} \,dx\wedge dt \,,
	\end{equation}
	Equivalently, 
	\begin{equation}\label{CurvForm}
	R(H) = \frac{1}{2} n_t\Gamma^{t}_{[x t]} \,dx\wedge dt =  (a_xn_t - a_tn_x)\,dt\wedge dx  = (a_xn_t)\,dx\wedge dt = R_{xt}(H)\, \,dx\wedge dt \,.
	\end{equation}
	Now we can see that in \textit{torsionless} NC geometry, $dn$ is a closed 1-form, i.e. $ dn=0 $ that corresponds to $ R_{\mu\nu}(H) = 0 $. This, in turn, translates to zero curvature in the gauge field i.e. a flat connection, corresponding to the time translation symmetry generated by the Hamiltonian. On the other hand, the TTNC case corresponds to a non-zero  $R_{\mu\nu}(H)$ or $ dn\neq0 $. The Frobenius condition then tells us that this curvature is given by the torsion tensor $a_{\mu}$: $ \partial_{[\mu}n_{\nu]} = a_{[\mu}n_{\nu]}$. Lifshitz field theories with classical Weyl invariance couple to TTNC geometry and the Weyl anomaly will be directly related to this torsion or acceleration vector field.
	
	To derive the anomalous Ward identity, we start with a classical action $S[{\phi}, N_i, h_{ij}]$ with matter fields ${\phi}$ coupled to background TTNC geometry and which is invariant under infinitesimal anisotropic local Weyl transformation with scaling exponent $z$
	\begin{equation}
	\delta N = z\sigma N, \quad h_{ij} = 2\sigma h_{ij} \,,
	\end{equation}
	where $\sigma(x,t)$ is the infinitesimal Weyl transformation parameter.
	Quantum mechanically, however the regularization of UV infinities of the partition function $Z = e^{-W[N,N_{i},h_{ij}]} = e^{-W[N,N_i,h_{ij}]}$ breaks the local Weyl invariance of the quantum effective action $W[N,h_{ij}]$ resulting in a Weyl anomaly. More concretely, the presence of a Weyl anomaly in the effective action necessarily means that 
	\begin{equation}
	\delta W = \int N \sqrt{h}\, dt\ dx \, \sigma\, \mathcal{A} \,,
	\end{equation}
	where quantum mechanically $ \mathcal{A} $ is given by the expectation values of the trace of the energy-momentum tensor
	\begin{equation}
	\mathcal{A} = z\mathcal{\langle E \rangle } + \langle \mathcal{P}^{i}_{i}\rangle = z\left\langle T^t_t \right\rangle +\left\langle  T^x_x  \right\rangle \neq 0 \,.
	\end{equation}
	It is important to note that although \cite{Oz2014} in their cohomological classification does not explicitly say that the background geometry to which they couple the Lifshitz theory is a TTNC spacetime, it actually implicitly is. For the cohomological classification of Weyl anomalies in FPD-invariant Lifshitz field theories in all spacetime dimensions, the foliation 1-form $n_{\mu}$ satisfies the Frobenius condition which is the key defining property of TTNC geometry. Section 2.4 of \cite{Oz2016} contains more information on the relationship between the notations and conventions used in \cite{Oz2014} and standard NC geometry. 
	
	We now move to demonstrate the geometric and physical nature of the Lifshitz Weyl anomaly after rewriting it in terms of the ADM coordinates defined above. We emphasize that the discussion in this section is valid for \textit{all} values of $z$. We will present two different yet related pictures. While the first stresses that $n_{\mu}$ and $n^{\mu}$ are the fundamental objects in the TTNC geometry, the second one stresses the key role of the torsion vector itself $ a_{\mu}$ in generating the Lifshitz Weyl anomaly. The latter picture will turn out to be useful in Section \ref{AnomalyDerivation} and Section \ref{Discussion} when the anomaly is derived from the (2+1)-dimensional NRSCS action.  
	
	\subsubsection{The 1-form Picture}
	In terms of the ADM gauge in (\ref{ADMTimelikeFol}) and (\ref{LieADM}), the Weyl anomaly is given by the variation of the one-loop effective action of the (1+1)-dimensional Lifshitz effective action $W[g]$
	\begin{eqnarray}\label{Weylanomaly}
	\delta W  &=& \int \sqrt{-g}\ \sigma \,  \tilde \epsilon^\mu \ \mathcal{L}_n a_\mu \\ \nonumber
	&=& \int \sqrt{-g}\ \sigma \, n_x\epsilon^{xt} \, \mathcal{L}_n a_t + n_t\epsilon^{tx} \, \mathcal{L}_n a_x \\ \nonumber
	&=& \int \sqrt{-g}\ \sigma \, n_t\, \mathcal{L}_n a_x \,,
	\end{eqnarray} 
	where $ \tilde{\epsilon^\mu} = n_\alpha \epsilon^{\alpha \mu} $ is the foliation-projected Levi-Civita tensor, $\epsilon^{tx}=1$, $ a_t = N^xa_x = 0 $, and $ \sigma(x,t) $ is the Weyl transformation parameter. Using the definition of the Lie derivative in (\ref{TorsionADM}), the Weyl anomaly is given by
	\begin{eqnarray}\label{AnomnalyTorsion}
	\delta W\, &=& -\int dt dx\ N^2 \sqrt{h}\, \sigma \, \mathcal{L}_na_x \\ \nonumber
	&=& -\int  dt dx\ N \sqrt{h}\, \sigma \, \left(\dfrac{\del a_x}{\del t} \right) \,.
	\end{eqnarray}
	In terms of the lapse function $ N(x,t)$, using (\ref{CurvForm}) and (\ref{TraceTorsionTensor}), it takes the following form
	\begin{eqnarray}\label{AnomalyLapse}
	\delta W &=& - \int dt dx\ N\sqrt{h}\, \sigma \, \left( \frac{1}{N}\partial_t\partial_x N  - \frac{1}{N^2} \del_t N\del_x N \right) \nonumber \\
	&=& - \int dt dx\ N\sqrt{h}\, \sigma \, \left( \frac{1}{N}\partial_t R_{xt}(H)  - \frac{1}{N} \partial_t N a_x\right) \nonumber \\
	&=& - \int dt dx \sqrt{h}\, \sigma \, \left(\partial_t R_{xt}(H)  - \partial_t N a_x\right) \,.
	\end{eqnarray}
	Expressing $\left\langle T^{\mu}_{\mu}(x)\right\rangle = \tilde \epsilon^\mu \ \mathcal{L}_n a_\mu $ in terms of local tangent frame coordinates and using differential forms will better reveal its geometric nature. Using that the vielbeins for the temporal and spatial components of the NC metric $g_{\mu \nu} = \gamma_{\mu \nu} - n_\mu n_\nu$ can be expressed as 
	\begin{equation} \label {ADMMetricVielbein}
	n_{\mu \nu} = n_{\mu} n_{\nu} \,, \quad  h_{\mu \nu} = e_{\mu}{}^{A}\delta_{AB}e_{\nu}{}^{B} \,,
	\end{equation}
	the vielbeins for the ADM metric in 1+1 dimensions in (\ref{ADMMetric}) are then given by
	\begin{equation}
	n=N dt, \qquad e^1 = (N^x dt + e^1{}_x dx )\,,
	\end{equation} 
	and the torsion coefficients are given by
	\begin{eqnarray}\label{TorCur}
	dn&=&\left(\del_x N \right)  dx\wedge dt = \frac{\del_1N}{N} e^1\wedge n \\ \nonumber
	de^1&=&\left(\frac{\del_xN^x}{2N} -\frac{e_1^x\del_t e_x^1}{2N}
	\right) e^1\wedge n = -\frac{K}{2}\ e^1\wedge n \,,
	\end{eqnarray}
	where $ K $ is the trace of the extrinsic curvature tensor defined in (\ref{ExtrCurvADM}).  We now use \textit{Cartan's formula} $ \mathcal{L}_X d\omega = d\mathcal{L}_X \omega $, which relates the Lie derivative along a vector field $ X $ of a $ k $-form $ d\omega $ to the exterior derivative of the $ (k-1)$-form $\Lie{X} \omega$. Acting with the Lie derivative on $ dn$ along $n^{\mu}$, we get
	\begin{eqnarray}\label{CartanFormula}
	\Lie{n} (dn) = d\Lie{n}n &=& d\Lie{n}\,(N\, dt) \\ \nonumber
	&=& d( \frac{1}{N}\del_t N - N^x \del_x a_x) dt \\ \nonumber 
	&=& (\del_t a_x) dt\wedge dx \,,
	\end{eqnarray}
	where we used the definition of the Lie derivative in (\ref{LieADM}) and (\ref{TorsionADM}) and chose $N^x$ to be zero. The expectation value of the energy-momentum tensor is therefore given by
	\begin{equation}\label{AnomTraceEM}
	z\left\langle T^t_t \right\rangle +\left\langle  T^x_x  \right\rangle = \del_t a_x = \del_t \left(\frac{R_{xt}(H)}{N(x,t)}\right)\,.
	\end{equation}
	From equations (\ref{CartanFormula}) and (\ref{AnomTraceEM}), we can see that the 1+1 Lifshitz Weyl anomaly, in the 1-form picture, is naturally given by the time derivative of the curvature of the timelike foliation 1-form $ n_\mu $ or equivalently as the time derivative of $a_x$, the solution of the Frobenius condition (\ref{CurvForm}). This makes explicit the relationship between TTNC geometry, the Frobenius condition and the role they both play in the generating the 1+1 Lifshitz Weyl anomaly.

\subsubsection{The 2-form Picture}\label{2-form-picture}
\begin{figure}
	\centering
	\includegraphics[width=0.7\linewidth]{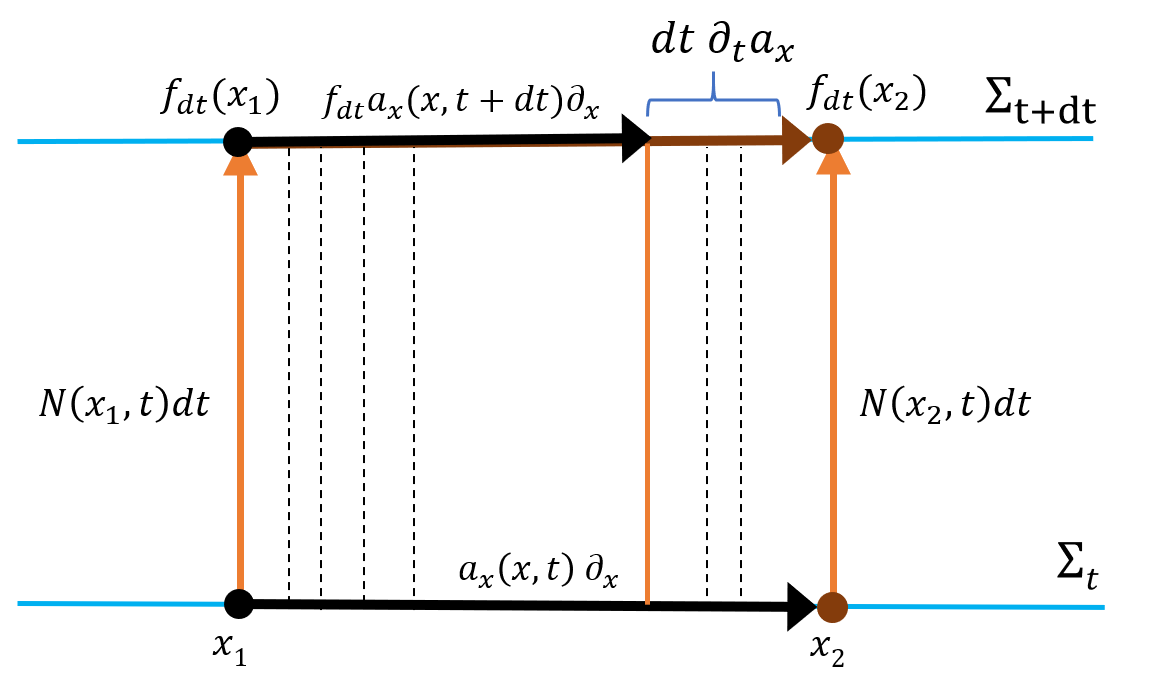}
	\caption{\textbf{Figure 1:} Geometrical depiction of the Weyl anomaly. Lie dragging the tangent vector $a_x(x,t) \del_x$ with end points $x_1$ and $x_2$ defined on a spatial slice $\Sigma_t$ at coordinate time $t$ to another slice $\Sigma_{t+dt}$ using the diffeomorphism $f_{dt}$ associated with the vector $n_{\mu} = Ndt$. The points $x_1$ and $x_2$ are Lie dragged to $f_{dt}(x_1)$ and $f_{dt}(x_2)$ respectively, while the vector $a_x(x,t) \del_x$ is dragged to $a_x(x,t+dt) \del_x$. The Lie derivative of the vector $a_x(x,t) \del_x$ along $n_{\mu}$ is then the difference between actual value of the vector at $f_{dt}$  and its value as a result of the Lie drag at $f_{dt}(x_2)$. The difference is the Weyl anomaly $dt\,\del_t a_x$. Notice that as a result of the Weyl anomaly, the vector $a_x$ on $\Sigma_t$ has been \textit{scaled} on $\Sigma_{t+dt}$ and therefore the spacetime area $(x_2 - x_1)dt$ has been \textit{squeezed}.}
	\label{fig:lie-drag2}
\end{figure}

	In the 2-form picture, the torsion vector is expressed as a 1-form
\begin{equation}\label{Tor-1form}
	a = a_\mu dx^{\mu} = a_t \, dt + a_x \, dx \,.
\end{equation}
The curvature 2-form $D_{\mu \nu}$ of the torsion 1-form $a_{\mu}$ is then given by 
\begin{equation} \label{Tor-curv}
	D = da = \left(\dfrac{\del a_t}{\del x} -\dfrac{\del a_x}{\del t} \right)\,dx\wedge dt \,.
\end{equation}
	Using $ a_{t} = N^{x}a_{x} $, $ da $ can be expressed as:
\begin{equation} \label{Lie}
	da = \left(\dfrac{\del a_x}{\del t} - N^{x}\dfrac{\del a_x}{\del x}\right)\,dt\wedge dx \,.
\end{equation}
	Setting $a_t = 0$ or equivalently, $ N^x =0 $, we get 
\begin{equation} \label{Curvature-2-form}
	da = \left(\dfrac{\del a_x}{\del t}\right)\,dt\wedge dx \,.
\end{equation}
which illustrates the 1+1 Lifshitz Weyl anomaly can be directly interpreted as the curvature of the torsion 1-form $ a_\mu$. We can clearly see that the anomalous \textit{gravitational} degree of freedom is a direct consequence of the \textit{time-dependence} of the lapse function $N(x,t)$. Hence, the presence of this 1+1 Lifshitz Weyl anomaly necessarily implies that energy is not conserved in the system. Equivalently, since the acceleration vector is time-dependent, the system experiences \textit{jerk}.  To get an intuitive interpretation of this Weyl anomaly and what it physically implies, we use the picture of Lie dragging in Fig. \ref{fig:lie-drag2}.  Expressed in this way, $a_\mu$ is the principal $G $-connection 1-form on a principal $G$-bundle $P$ over a smooth manifold $M$ with values in the Lie algebra $ \mathfrak{g} $ of $G$ while $D_{\mu \nu}$ is its curvature form. The associated vector bundle in our case, is the \textit{dual} or cotangent frame  bundle over $M$.  In Section 3.3, we will see that $ G = SO^+(1,1)$, the identity component of the indefinite orthogonal group $SO(1,1)$ of area-preserving \textit{squeeze} transformations which also happens to be the restricted group of Lorentz boosts in 1+1 dimensions.

As we will also point out in Section \ref{AnomalyDerivation}, in the process of gauging the Schr$\mathrm{\ddot{o}}$dinger algebra in 2+1 dimensions, the torsion 1-form $a_{\mu}$ appears as the gauge connection of the Weyl symmetry and thus it is natural to expect that the Weyl anomaly would involve the curvature $da$.

\section{Derivation of the Weyl Anomaly From the NRSCS Action} \label{AnomalyDerivation}
In this section, we derive the $z=2$ Lifshitz Weyl anomaly from a 2+1-dimensional (3D) non-relativistic Schr$\mathrm{\ddot{o}}$dinger-invariant Chern-Simons action on a manifold with a boundary. The boundary theory is a $z=2$ Lifshitz theory coupled to TTNC geometry. This 3D NRSCS action was recently constructed by gauging the centrally-extended Schr$\mathrm{\ddot{o}}$dinger algebra which made dynamical the TTNC geometry. In the metric formalism, it was then shown that the NRSCS action is indeed equivalent to a three-dimensional non-projectable $z=2$ \textit{conformal} or Weyl-invariant HL theory of gravity which is the counterpart of relativistic conformal gravity. As we will discuss below, this NRSCS action contains two terms which do not contribute to the solution of the bulk equation of motion. In this section, we place the 3D NRSCS action on a manifold with \textit{boundary} and show that one of these two terms, the tCS term, does in fact generate the $z=2$ (1+1)-dimensional Lifshitz Weyl anomaly in (\ref{Weylanomaly}). In fact, the authors in \cite{ObersNRSCS} wondered if one of these two terms would correspond to a boundary Weyl anomaly. Let us emphasize again that the (1+1)-dimensional Lifshitz Weyl anomaly we are discussing in this paper is \textit{universal} i.e. true for \textit{all} values of $z$. Therefore, throughout the discussion in this section, the relevant dual boundary theory is a $z=2$ Lifshitz theory with a background TTNC geometry. 
	
\subsection{Key Proprties of the NRSCS Action}
By gauging the \textit{extended} Schordinger algebra, i.e. by letting the gauge field $ A $ take its value in the centrally-extended Schr$\mathrm{\ddot{o}}$dinger algebra, $A$ can be expanded as a linear combination of the generators of the Schr$\mathrm{\ddot{o}}$dinger group \cite{ObersNRSCS}
\begin{eqnarray} 
A & = & H \tau +P_a e^a +G_a \omega^a + J \omega + N m +W a +K f \nonumber\\
&&+S\zeta+Y \alpha  + Z\beta \,, \label{eq:ASch}
\end{eqnarray} \label{NRSCS}where $H$, $P_a$, $G_a$, $J$, $N$, $W$, $K$ are the generators of time translations (the Hamiltonian), momentum translation, Galilean boosts, rotations, central element, Weyl transformations and special conformal transformations, respectively.\footnote{The generator of scale or dilatation transformation in \cite{ObersNRSCS} is denoted by $D$ which we reserve here for the curvature of the $a_\mu$ connection.} The three central extensions of the Schr$\mathrm{\ddot{o}}$dinger group are $S$, $Y$ and $Z$ respectively. Using the metric on this non semi-simple Lie algebra, the NRSCS action as given in \cite{ObersNRSCS} is
\begin{eqnarray}
S_{NRSCS} [A] &=& \frac{k}{4\pi} \int_M Tr \left[A \wedge dA + \frac{2}{3} A \wedge A \wedge A \right] \label{action3} \nonumber \\
&=& \frac{k}{2\pi} \int_M Tr\ c_1 \mathcal{L}_{NRSCS} {+c_2\left[a\wedge da-\tau\wedge df+2a\wedge\tau\wedge f\right]}+2c_3\omega\wedge d\omega \\
&=& \frac{k}{2\pi} \int_M Tr\ c_1 \mathcal{L}_{NRSCS} +c_2\,\mathcal{L}_{tCS} +2c_3\omega\wedge d\omega\,,
\end{eqnarray}
which for $ c_2 = c_3 = 0 $ is equivalent to a bulk 3D action of non-projectable conformal HL gravity. The arbitrary constants $ c_i $ are defined in terms of the symmetric bilinear form invariant under the Schr$\mathrm{\ddot{o}}$dinger algebra, i.e. $ B(V_i, V_j) = c_i $ and $ V_i $ is a generator of the algebra. For example, $ B(D,D) = 2c_2 $. A key observation is that $a_{\mu}$ is the gauge field of the dilatation symmetry. The curvature of the torsion vector is given by $R_{\mu \nu}(D) = da - 2df$ where $f_{\mu}$ is the gauge field associated with the generator of special conformal transformations $K$. (In this section $K$ is not the trace of the extrinsic curvature). Therefore, one should expect that a boundary Weyl anomaly would be generated by the tCS term in the action.
	
With $ c_{2} = c_{3} = 0 $, the NRSCS action (\ref{NRSCS}), satisfies a bulk equation of motion $F=dA+A\wedge A=0$ that gives a $ z=2 $ Lifshitz metric $ ds^2=-\frac{dt^2}{r^4}+\frac{dr^2}{r^2}+\frac{dx^2}{r^2}\,$. Therefore, the bulk theory represented by the NRSCS action is \textit{Weyl-invariant}. However, the tCS term whose coefficient is $c_2$, transforms under the $SL(2,\mathbb{R}) $ subgroup of the Schr$\mathrm{\ddot{o}}$dinger algebra and as discussed in \cite{ObersNRSCS} cannot be removed by a field redefinition and therefore, as noted in \cite{ObersNRSCS}, it may lead to a Weyl anomaly at the boundary. The tCS term is an Abelian CS term composed entirely of $a_{\mu}$, the gauge connection of the local Weyl symmetry and therefore, under a Weyl transformation on a manifold with a boundary, it is natural to expect that the total derivative term leads to a Weyl anomaly of the boundary effective action. In other words, the sole contribution of the tCS term when added to an on-shell bulk HL theory of gravity is to generate the Weyl anomaly of the \textit{dual} boundary theory since, as discussed above, it does \textit{not} contribute to the solution of the bulk $z=2$ HL gravity theory. It may be worth pointing out that the $SL(2,\mathbb{R})$ is isomorphic to $SO^+(2,1)$ which is the group of Lorentz transformation in three dimensions. At the boundary, where there is a Weyl anomaly, the torsion 1-form $a_{\mu}$ transforms under $SO^+(1,1)$ as we explained in the previous section. Recently, it was shown that the NRSCS action can be reformulated into a manifestly 3D relativistic form due to the presence of the $SL(2,R)$ subalgebra in the extended Schroedinger algebra \cite{Sorokin19} and therefore can be interpreted as a relativistic CS theory.
	
\subsection{The Lifshitz Weyl Anomaly from the tCS Term}\label{AnomalyFromtCS}
Denote the (2+1) on-shell HL gravity action by $S_{HL}$. Using the developed machinery of non-relativistic holography \cite{Thompson2012, KarchNRHolo}, which started when Lifshitz and Schr$\mathrm{\ddot{o}}$dinger spacetime solutions to relativistic actions of gravity were found \cite{Son08, McGreevy08, Kachru08, Taylor08}, the variation of the on-shell HL action at low energies and to leading order in the metric can be expressed in terms of the TTNC geometry on the boundary
\begin{equation}\label {NRHolog}
	\delta S_{HL} = \int d^3x \sqrt{g^{(0)}} T^{ij}g_{ij}^{(0)} \,,
\end{equation}
where $d^3x \equiv dt dx dr$ and $\sqrt{g^{(0)}} = N^{(0)}\sqrt{h^{(0)}}$ is the metric in terms of the boundary lapse and shift vectors and $T^{ij}g_{ij}$ is identified with the trace of the expectation value of the boundary theory effective action $W[N^{(0)},h_{ij}^{(0)}]$ coupled to a metric that is \textit{anisotropically} conformal to $N^{(0)}h_{ij}^{(0)}$.\footnote{To properly define an asymptotically Lifshitz spacetime, we assume the notion of anisotropic conformal infinity of the $D+1$-dimensional Lifshitz geometry at $r\rightarrow 0$ where there is an asymptotic codimension-one foliation \cite{Thompson10}.} If the action $S_{HL}$ is Weyl-invariant as for example the one constructed in \cite{ObersNRSCS}, then $\delta S_{HL} = 0$ under the variations $\delta N = z\sigma N^{(0)} \, \, \mathrm{and} \, \, h_{ij}^{(0)} = 2\sigma h_{ij}^{(0)}$. However, if we were to trust the machinery of non-relativistic holography especially for HL gravity and asymptotically Lifshitz spacetimes, we have to be able to deal with a Weyl-anomalous boundary theory and assume a non Weyl-invariant bulk theory of gravity with a non-vanishing  $T^{ij}g_{ij}^{(0)} = 2\left\langle T^t_t \right\rangle +\left\langle  T^x_x  \right\rangle $. Adding the tCS term to the on-shell gravity action $S_{HL}$ is our way out. As we show below, under a Weyl transformation, the tCS term is invariant up to a boundary term. If we assume the coefficient of the tCS term matches that of the boundary Weyl anomaly, then it cancels with the variation of the bulk on-shell action. More concretely, the variation of the on-shell bulk gravity action $S= S_{HL} + S_{tCS}$ under a Weyl transformation with parameter $\sigma(x,t)$ should be given by 
\begin{equation}\label{Var_On-shell_Gravity_Action}
	\delta_{\sigma} S =\delta_{\sigma} S_{tCS} = \frac{k}{2\pi} c_2 \int_{\del M} Tr [\delta a\, da] \,.
\end{equation}
Let us see how we do that. We set $c_3 = 0$ in the NRSCS action and start by integrating out the connection $ \beta$ in NRSCS action. The corresponding equation of motion is  $ df = -2b\wedge f$. Substituting this solution into the tCS term, we see that $- \tau \wedge df $ cancels with $ 2b\wedge \tau\wedge f $ such that the tCS term can be written as 
	\begin{equation}\label{tCSAction}
	\mathcal{L}_{tCS}[a] = 2c_{2} \left(a\wedge da \right) \,.
	\end{equation}
	In terms of differential forms, a variation of the torsion field $ a $  in $ S_{tCS}[a] $ gives
	\begin{equation}\label{TorsionalActionVar}
	\delta S_{tCS}[a] = \frac{k}{2\pi}c_2 \int_M \, Tr[\delta a\,\wedge D] + \int_{\del M} Tr [\delta a\, da]\,,
	\end{equation}
	where $k$ is the level of the NRSCS action.
	In components with coordinates $(t,x,r)$, the $S_{tCS}[a]$ action reads
	\begin{equation}
	S_{tCS}[a] = \frac{k}{2\pi} c_2 \int d^3x\ \epsilon^{\mu\nu\rho}a_\mu \partial_{\nu}a_{\rho}\label{tCS} \,.
	\end{equation}
	The variation of the tCS action is then given by
	\begin{eqnarray} 
	\delta S_{tCS}[a]  &=& \frac{k}{2\pi} c_2\int d^3x\ \ep \left[ \delta a_\mu\partial_\nu a_\rho +  a_\mu \partial_\nu \delta a_\rho\right] \nonumber\\  
	&=& \frac{k}{2\pi} c_2\int d^3x\ \ep \left[ \delta a_\mu D_{\nu\rho} +  \partial_\mu (a_\nu \delta a_\rho)\right]
	\end{eqnarray}
	$D_{\mu\nu}=0$ is the equation of motion that minimizes the tCS action. The last term must be set to zero on the boundary at $r=0$. One choice is  
	\begin{equation}\label{tCSBoundaryGaugeChoice1}
	(a_t - N^x a_x) \Big|_{r=0} = 0, \quad  N^r =0 \,,
	\end{equation}
	since by definition, $ a_t =  N^x a_x +  N^r a_r $. The other sets only $a_t$ to zero
	\begin{equation}\label{tCSBoundaryGaugeChoice2}
	a_t = N^x a_x = 0 \,.
	\end{equation}
	The choice in (\ref{tCSBoundaryGaugeChoice1}) is however more general. Under infinitesimal local Weyl transformation with parameter $\sigma(x,t)$, the gauge connection $a_{\mu}$ transforms as 
	\begin{equation}
	\delta_{\sigma}a = d\sigma \,,
	\end{equation}
	and the tCS action varies by a the total derivative term
	\begin{equation}
	\delta_{\sigma} S_{tCS}[a] = \frac{k}{2\pi} c_2\int_{\del M} Tr (\sigma da)
	\end{equation}
	In components, this becomes 
	\begin{equation}
	a_\mu \rightarrow a_\mu + \partial_\mu \sigma \nonumber 
	\end{equation}
	and 
	\begin{equation}
	S_{tCS}\rightarrow S_{tCS} + \frac{k}{2\pi} c_2\int_{r=0} dxdt\ \sigma (\partial_t a_x - \partial_x a_t)\nonumber \,,
	\end{equation}
	which has precisely the same \textit{form} of the (1+1)-dimensional Weyl anomaly in (\ref{AnomnalyTorsion}) and (\ref{Tor-curv}) of the Lifshitz boundary theory. Thus, the bulk remains Weyl-invariant while the boundary theory does not. We can then conclude that the tCS term added to a 3D Weyl-invariant HL gravity action plays a role similar to what the gCS term plays when added to a 3D diffeomorphism-invariant action. This is the main result of this paper. However, it is important to observe that without knowing the exact value of the coefficient $c_2$ and matching it with that of the anomaly computed in an example Lifshitz field theory, for example, using heat kernel methods, it would be difficult to claim the derivation is exact.
	
	It stands to reason that we should be able to find the $ a\wedge da $ term in the parity-odd sector of the cohomology of the relative Weyl operator in 2+1 dimensions. Indeed, we found such a term in \cite{Oz2014}. The term is given by
	\begin{eqnarray} \label{OztCS}
	\tilde{\epsilon}^{\alpha \beta} a_{\alpha}\Lie{n}a_\beta &=& \tilde{\epsilon}^{x r} a_{x}\Lie{n}a_r + \tilde{\epsilon}^{r x} a_{r}\Lie{n}a_x  \nonumber \\
	&=& a_{x} \del_{t} a_r - a_{r} \del_{t} a_x  \,,
	\end{eqnarray}
	where we have used that $ \Lie{n} a_{r} = \frac{1}{N} \left(\del_{t} a_r - N^r \Lie{N^r} a_{r} \right) $, $\tilde{\epsilon}^{\alpha \beta} = n_{\gamma}\epsilon^{\gamma \alpha \beta} $ and $a_t = 0 $. Now let us show that $  a\wedge da $ can be expressed as (\ref{OztCS}). 
	Let us start by expanding the $  a\wedge da $ in coordinate bases as $ a = a_\mu\,dx^\mu = a_x\, dx + a_r\, dr + a_t\, dt $. The $  a\wedge da $ term can be expanded as follows
	\begin{equation}
	\begin{split}
	a\wedge da &= \epsilon^{ijk} a_i\del_j a_k \nonumber 
	\\
	&
	= a_t\left( \del_x a_r - \del_r a_x  \right) dt dx dr - a_x\left( \del_t a_r - \del_r a_t  \right) dx dr dt 
	\\
	&
	+ a_r\left( \del_t a_x - \del_x a_t  \right) dr dx dt \,,
	\end{split}
	\end{equation} 
	where $di\, dj\, dk\, \equiv di \wedge dj \wedge dk$. The exterior derivative $ da $ is given by
	\begin{eqnarray}
	a\wedge da  &=& \epsilon^{xr}a_x\del_ta_r \nonumber \\
	&=& a_x\del_ta_y - a_r\del_ta_x 
	\end{eqnarray}
	which matches the one given in (\ref{OztCS}). 
	
	\section{The $z=1$ Lifshitz Weyl Anomaly}
	In this section, we focus on the interesting case of the $z=1$ Lifshitz Weyl anomaly. We will illustrate that $\left\langle T^\mu_{(e)\mu}\right\rangle$ can be interpreted as the scalar curvature of the \textit{dual} Lorentz connection 1-form in a two dimensional local gravity theory constructed in \cite{Solod90} with conformal and Lorentz anomalies.
	\subsection{The $z=1$ Lifshitz Weyl Anomaly as the Scalar Curvature of the Dual Lorentz Connection} \label{Weyl-Lorentz-Anom-Relation}
	The authors in \cite{Oz2014} revealed that the $z=1$ (1+1)-dimensional Lifshitz Weyl anomaly is the \textit{Weyl partner} of the Lorentz anomaly in 1+1 CFT. The idea was to shift the Lorentz anomaly of a $1+1$-dimensional CFT to a foliation dependence, i.e. to a \textit{dependence} on $ n_\mu $ and rewrite the \textit{anomalous} CFT quantum effective action in terms of $ n_\mu$ as follows 
	\begin{equation}\label{compartoconf:lifactiondef}
	W_{\text{Lif}}[e^a{}_\mu,t^a] \equiv W_{\text{conf}} [-e^a{}_\mu n_a, e^a{}_\mu \widetilde{n}_a] = W_{\text{conf}}[-n_\mu, \widetilde{n}_\mu],
	\end{equation}
	where $n_{\nu}$ and $\widetilde{n}^a \equiv  \epsilon^{ab} n_b$ are arbitrary foliation vectors aligned with the frame fields $e^a{}_\mu$ which are defined for a relativistic spacetime as
	\begin{align}
	g^{\mu \nu} e^{a}_{\mu} e^{b}_{\nu} &= \eta^{ab}  \\
	\eta_{ab} e^{a}_{\mu} e^{b}_{\nu} &= g_{\mu \nu} \,,
	\end{align}
	where $\eta^{ab}$ is the flat metric in the two-dimensional tangent frame basis. The Lorentz spin connection, in terms of $ e^{a}_{\mu} $,  is defined as
	\begin{equation}\label{SpinConnection}
	\omega_\mu{}^a{}_b = - e_b{}^\nu \nabla_\mu e^a{}_\nu \,.
	\end{equation}
	and the connection 1-form $ \omega_\mu $ is given by
	\begin{equation}
	\omega_{\mu} = \epsilon^{ab}\omega_{ab,\mu} = \epsilon^{ab}\partial_{\mu}\omega_{ab}
	\end{equation}
	It is well known \cite{BertlandBook} that in a 1+1 CFT with local Lorentz anomaly \footnote{A diffeomorphism anomaly in 1+1 CFT can be shifted to a local frame anomaly by a local counterterm.}, for example, in a chiral CFT, the Weyl anomaly has an extra term in addition to the Ricci scalar $ R $. This extra term is essentially how the Lorentz anomaly \textit{manifests} itself in $\left\langle T^\mu_{(e)\mu}\right\rangle$ taken as the variation of $ W_{\text{Lif}}[e^a{}_\mu]$ with respect to the viebein 1-forms $e^a{}_\mu $. This additional term is a total divergence of the Lorentz (spin) connection 1-forms  $\omega^\mu{}_{ab}$ defined in (\ref{SpinConnection})
	\begin{equation}\label{LorentzWeylPartner}
	\left\langle T^\mu_{(e)\mu}\right\rangle   = -2\epsilon^{ab} \nabla_\mu \omega^\mu{}_{ab} \,.
	\end{equation}
	After identifying local tangent frame i.e. the vielbeins with the foliation 1-forms
	\begin{equation} \label{vielbein-Iden}
	e^{0}_{\mu} \equiv n_{\mu}, \quad e^{1}_{\mu} \equiv \widetilde{n}_\mu \,,
	\end{equation}
	the authors in \cite{Oz2014}, were able to demonstrate that $\left\langle T^\mu_{(e)\mu}\right\rangle$ is indeed the Weyl partner of the Lorentz anomaly up to the coboundary terms $(a_\rho K + \wn_\rho K)$
	\begin{empheq}{align}
	\begin{split}
	\left\langle T^\mu_{(e)\mu} \right\rangle 
	= &\ -2 a \epsilon^{ab} \nabla_\mu \omega^\mu{}_{ab}
	\\ 
	&= -2a \nabla_\mu \left(\epsilon^{ab} e_{a\nu} \label{TraceEMLorentz} \nabla^{\mu}e_{b}{}^{\nu} \right)   \\
	& = 4a \tilde \epsilon^\rho (\Lie{n} a_\rho + a_\rho K + \wn_\rho K) \,,
	\end{split}
	\end{empheq}
	where $\wn_{\mu}$ is the foliation-projected covariant derivative of a foliation-tangent tensor $V_{\alpha \beta}$.\footnote{See equation 2.35 in \cite{Oz2014} for a definition of the foliation-projected covariant derivative.} This identification essentially maps the Poicare-invariant tangent vector space of the relativistic spacetime manifold to Schr$\mathrm{\ddot{o}}$dinger-invariant (or Bargmann-invariant) tangent vector space of the TTNC spacetime manifold to which Lifshitz field theories naturally couple. The original CFT effective action $W_{\text{conf}}[e^a_{\mu}]$ is technically no longer Lorentz-anomalous but rather foliation-dependent and hence the physical content of $\left\langle T^\mu_{(e)\mu}\right\rangle$ now reflects a Weyl anomaly rather than a Lorentz anomaly. In other owrds, the anomalous local frame rotations of $W_{\text{conf}}[e^a_{\mu}]$ is exchanged for anomalous Weyl transformations of the foliation in $W_{\text{conf}}[-n_\mu, \widetilde{n}_\mu]$.
	
	We now illustrate that the $\left\langle T^\mu_{(e)\mu}\right\rangle$ is related to the scalar curvature of the dual Lorentz connection 1-form of the local gravity action in \cite{Solod90}. In \cite{Solod90}, a \textit{local} action of two-dimensional gravity was constructed out of the frame fields $e^{a}_{\mu}$
	\begin{eqnarray}\label{SolodukhinAction}
	W[e^{a}_{\mu}] &=& \frac{1}{4} \int d^2 x \,e\, T^{a}{}_{\mu\nu}T^{\mu\nu}{}_{a} \\ \nonumber 
	&=& \frac{1}{4} \int d^2 x \,e \, (\del_{\mu}e^{a}{}_{\nu} - \del_{\nu}e^{a}{}_{\mu})\, (\del^{\mu}e_{a}{}^{\nu} - \del^{\nu}e_{a}{}^{\mu}) \,,
	\end{eqnarray}
	where $e$ = det $e^{a}{}_{\mu}$. Under a local conformal transformation $\delta e^{a}{}_{\mu} = \sigma e^{a}{}_{\mu}$ with infinitesimal Weyl parameter $\sigma$, the action suffers a conformal anomaly $R$ while under a local Lorentz transformation $\delta e^{a}{}_{\mu} = \chi \epsilon^{a}{}_{b} e^{b}{}_{\mu}$ with infinitesimal Lorentz parameter $\chi$, it has a Lorentz anomaly $U$. Using (\ref{SpinConnection}) and the fact that the two-dimensional Levi-Civita tensor obeys $\epsilon_{ab} + \epsilon_{ba} = 0$, the Ricci scalar $R$ can be expressed in terms of the curvature 2-form of the Lorentz connection $\omega$ 
	\begin{equation}
	\mathcal{R} = d\omega \,,
	\end{equation}
	while the scalar $U$ of the Lorentz anomaly can be expressed in terms of the curvature of the dual Lorentz connection $\star\omega$
	\begin{equation}
	\mathcal{U} = d\star\omega \,,
	\end{equation}
	where $\star$ is the Hodge dual operator. If the Lorentz connection 1-form is expressed in terms of $a_x$ and $K$ as 
	\begin{equation}
	\omega = a_x dt + \frac{K}{N} dx
	\end{equation}
	then $\int d\omega$ gives the foliation-projected decomposition of $R$ (not to be confused with the $R(H)$ of the Weyl anomaly)
	\begin{equation}\label{RicciFoliation}
	R = \int \mathcal{R} = \int dt dx\, N\sqrt{h} \left( K^2 + \frac{1}{N}\del_tK + \left( \frac{\del_xN}{N}\right)^2  - \frac{\del_x^2N}{N}\right) \,.
	\end{equation} 
	Similarly, if we define the dual Lorentz connection (which interestingly enough, only in two dimensions is also a 1-form) as
	\begin{equation}
	\star \omega = a_x dx - \frac{K}{N} dt \,,
	\end{equation}
	then 
	\begin{equation}\label{DualLorentzConn}
	d\star \omega  = \left(  \del_t a_x +  \frac{1}{N} \left(\del_x K - a_xK \right)\right)  dt\wedge dx \,,
	\end{equation}
	and hence $\left\langle T^\mu_{(e)\mu}\right\rangle$ can be expressed as the scalar curvature 2-form $d\star \omega$
	\begin{equation} \label{LorentzScalar}
	U\equiv \left\langle T^\mu_{(e)\mu}\right\rangle = 2 \nabla_{\mu}{}(\omega^{\mu}) = 2\nabla_{\mu}{}\left(\epsilon_{ab}e^{a}{}_{\nu}T^{b\mu\nu} \right) \,. 
	\end{equation}
	By comparing equations (\ref{LorentzScalar}) with (\ref{TraceEMLorentz}), we see that they have precisely the same form when decomposed in terms of foliation geometry. Therefore, we see that the 1+1 Lifshitz Weyl anomaly $\tilde \epsilon^\rho \Lie{n} a_\rho$ is directly related to the curvature scalar  $d\star \omega$ of the dual Lorentz connection (modulo the coboundary terms in (\ref{TraceEMLorentz}) and (\ref{DualLorentzConn})) when expressed in terms of the foliation 1-forms in (\ref{vielbein-Iden}). We note that the same curvature scalar $U$ of $\star \omega$ also appears in the process of quantizing a \textit{non-local} chiral gravity action which is known to have local Weyl as well as Lorentz anomalies \cite{Myers92}. It was also pointed out in \cite{Myers92} that the gauge group defined on the dual frame bundle related to $\star \omega$ connection 1-form is multiplication by positive real numbers which is consistent with the Lie drag picture in Fig. \ref{fig:lie-drag2}.
	
	More importantly, it was pointed out in \cite{Solod90} and \cite{Myers92} that $U$ like $R$ is also topological invariant i.e.
	\begin{equation} \label{IntegratedLifAnomaly}
	\lambda = \int_{M} \, e\, d^2x \,U + \int_{\del M} \,\sqrt{h\,} d\tau \,\mathrm{(boundary \, \, term)}  =  \mathrm{topological \, \, invariant} \,,
	\end{equation}
	where $e$ is det $e^a{}_{\mu}$. In fact, $U$ was obtained from the index of a generalized Dirac operator (see equation 28 in \cite{Solod90} and by using the conformal-Lorentz gauge, $\lambda$ was expressed as the boundary integral of the divergence of a unit tangent vector in the conformal-Lorentz gauge (here tangent to the \textit{spatial} foliation leaf)
	\begin{equation}\label{AnomalyTopo}
	\lambda = \int_{\del M}\, d\tau\, \nabla_{\mu} \hat{V}^{\mu} \,.
	\end{equation}
	The physical meaning of $\lambda$ as a conserved boundary charge is still not clear but we will discuss this observation further in Section \ref{Discussion} in light of recent progress in constructing boundary conformal invariants of type-B anomalies.
	
	\subsection{Weyl Anomaly Cancellation in $ z=1 $ Lifshitz Effective Action}\label{AnomalyCancellation}
	In this section, we discuss how canceling the Weyl anomaly leads to an equations of motion for a stationary chiral boson, which by solving, we obtain the Rindler metric. We then use Darboux's coordinates to explain how as a result of canceling the Weyl anomaly, the cotangent bundle of the spacetime manifold is a symplectic manifold with a \textit{Hamiltonian} function.
	
	To restore the local Weyl symmetry of the induced effective action, the Weyl anomaly must be canceled such that $a_x$ becomes conserved. Insisting on the Weyl invariance of the quantum effective action $W[e^a_{\mu}]$, amounts to satisfying the equation of motion in (\ref{AnomnalyTorsion}) or (\ref{AnomalyLapse}) by putting appropriate boundary conditions on $\del_tN$ at the spatial boundaries. The Weyl anomaly in (\ref{Weylanomaly}) and (\ref{AnomnalyTorsion}) assumes a zero shift vector $N^x = 0$. If $N^x = 0$, then the equation of motion is simply given by
	\begin{equation}\label{EOMZeroShift}
	\partial_t\partial_x N = 0 \,.
	\end{equation}
	Since, physically, the Weyl anomaly represents a \textit{time-dependent} acceleration, or a non-uniform gravitational field where energy is not conserved, restoring local Weyl invariance in the effective action is tantamount to having observers with constant, i.e. uniform proper acceleration in \textit{flat} spacetime or having a uniform gravitational field where energy is conserved. This necessarily means getting rid of the time dependence of the lapse function $N(x,t)$. Mathematically speaking, restoring local Weyl symmetry requires making $a_{\mu}$ a closed 1-form, i.e. a flat connection $da=0$ with zero curvature $D_{\mu \nu} = 0$.
	
	A natural question to ask is what the implications are of having a time-independent lapse function, one that only depends on the spatial coordinates $N(x)$. After all, in a projectable HL gravity theory, the lapse function is either only time-dependent $N(t)$ or time and spatially-dependent $N(x,t)$ in the non-projectable version. So, what does it mean to have time-independent lapse function as a result of canceling of the $z=1$ Lifshitz Weyl anomaly? We will comment on this peculiarity at the end of this section. 
	
	The equation of motion in (\ref{EOMZeroShift}) is that of a \textit{stationary} 1+1 gravitational \textit{chiral} boson whose general solution of (\ref{EOMZeroShift}) is given by
	\begin{equation}
	N(x,t) = N_1(x) + N_2(t) \,.
	\end{equation}
	Imposing the boundary condition  $\del_t N = 0$ at the spatial boundaries i.e. the spatial leaves of the foliation, necessarily means eliminating the $N_2(t)$ gauge degree of freedom and therefore (\ref{EOMZeroShift}) is automatically satisfied. By putting a boundary condition that sets $\del_t N$ to zero at the spatial boundaries, the Weyl anomaly is canceled and as a result, $a_x$ becomes the conserved charge of the local Weyl gauge symmetry. A spatially-dependent lapse function $N(x)$ then gives a family of arbitrary time-independent solutions each of which on a hypersurface of constant time $t$. Choosing $ N(x) $ to be linear is a particularly important choice of coordinates, since with this choice and $h_{xx}=1$, the background spacetime metric in ADM coordinates becomes
	\begin{equation}\label{RindlerMetric}
	ds^2 = -(\alpha x)^2\,dt^2 + \, dx^2 \,.
	\end{equation}
	which is the \textit{Rindler} metric of a hyperbolically accelerated reference frame with coordinates $(x,t)$  with rapidity $\eta=\alpha t$. If we label the flat Minkowski spacetime coordinates by $(X,T)$ and choose Rindler observer with constant proper acceleration $\alpha =1$ and proper time $\tau$ equal to coordinate time $t$, then $(X,T)$ are related to Rindler coordinates by the following transformations 
	\begin{equation}\label{RindlerWorldline}
	T=x \, \mathrm{sinh(t)}, \quad X = x\,\mathrm{ cosh(t)} \,.
	\end{equation}
	These linear transformations preserve the hyperbolae $X^2 - T^2 = N^2(x) = x^2$  which describe the worldlines of a family of Rindler observers at rest for \textit{fixed} $x$. These transformations can be represented by elements of the one-parameter group of Lorentz boosts $SO^+(1,1)$ with boost parameter $\eta=\alpha t$. An element in $SO^+(1,1)$ is represented by a $2\times2$ real matrix
	\begin{equation}\label{SOMatrixElement}
	M(\eta)=
	\begin{bmatrix}
	\mathrm{cosh}(\eta)	&  \mathrm{sinh}(\eta) \\ 
	\mathrm{sinh}(\eta)	& \mathrm{cosh}(\eta)
	\end{bmatrix} \,.
	\end{equation}
	In light-cone coordinates, $U=X+T, \ V=X-T$, $ M(\eta) $ is diagonalized
	\begin{equation}\label{SODiagonalMatrixElement}
	M(\eta)=
	\begin{bmatrix}
	e^{\eta}	&  0 \\ 
	0	& e^{-\eta} 
	\end{bmatrix} \,,
	\end{equation} 
	such that area $U*V = X^2 - T^2$ of the hyperbola is preserved. Therefore, the group $SO^+(1,1)$, in addition to being the group of Lorentz boosts in 1+1 dimensions is also the group of scale (actually squeeze) transformations that preserve the area $U*V$ of the hyperbolic worldline of a Rindler observer at a fixed $x=x_0$. 
	\\~\\
	If we define a frame fields $e^0$ and $e^1$ as
	\begin{equation}\label{Lapse-redshift}
	e^0 = x\,dt, \quad e^1 = dx \,,
	\end{equation} 
	which in terms of the dual basis vector field, is given by
	\begin{equation}\label{KillingVector}
	n^t = \frac{1}{x}\,\del_t, \quad n^x = \del_x \,,
	\end{equation} 
	then the unit timelike vector $n^{\mu}$ defines integral curves consisting of the world lines of a family of Rindler observers each at fixed $x=x_0$. For each such observer, $n^t$ is a \textit{Killing} vector of the Rindler metric which, when expressed in Minkowski coordinates, becomes the generator of Lorentz boosts in the $X$-direction
	\begin{equation}\label{BoostKilling}
	\del_t = (\del_tT)\del_T + (\del_tX)\del_X = X\del_T + T\del_X \,.
	\end{equation}
	Since the Lie derivative of the torsion vector $a_{\mu}$ along $n^{\mu}$ after canceling the Weyl anomaly is now $\Lie{n}a_{\mu} = 0$, $a_{\mu}$ is conserved and $n_{\mu}$ satisfies the Frobenius condition $n\wedge dn = 0$. It is interesting to note that the \textit{vorticity-free} condition of the  worldlines of Rindler observers i.e. the vanishing of the rotation tensor in the Raychaudhuri equation, is the twistless torsion condition in equation (6.8) of \cite{ObersDynmicTNC2015}.
	
	Some comments are in place. First, we observe that by making the lapse function time-independent, we have eliminated the extra foliation degree of freedom in HL gravity theories which as a result become Weyl-invariant \cite{Blas2011}. To understand the consequence of canceling the Weyl anomaly a bit further, we use \textit{Darboux's theorem} to show that the cotangent bundle of the flat spacetime is a symplectic manifold with a \textit{Hamiltonian} function $N(x) = x$ of a Hamiltonian vector field $n^{\mu}$ which is we what should expect anyway. Starting from the Frobenius condition, one can define local Darboux coordinates on a two-dimensional manifold $M$ with a 1-form $n_{\mu}$
	\begin{equation}
	n=x\, dt \,.
	\end{equation}
	Taking the exterior derivative of $n$ gives the \textit{symplectic} 2-from $R(H)$ on $M$
	\begin{equation}
	R(H)=dn= dx\wedge dt \,.
	\end{equation}
	since $R_{xt}(H)=\del_x N = 1$ by definition. Furthermore, using Cartan's formula, one can formally show that the Lie derivative of $R(H)$ vanishes
	\begin{equation}
	\Lie{n}R(H) = 0 \Leftrightarrow d(\iota_n R(H)) + \iota_n dR(H) = d(d\,N)+ dR(H) = 0 \,,
	\end{equation}
	where $\iota_n$ is the interior product, $dR(H)=0$ is automatically satisfied on a 2-dimensional manifold, $\iota_n R(H) = dN = dx$,  $d(dN(n^{\mu}))$ and $d(dx(\frac{1}{x}\del_t)) =0$ essentially means that $R(n^{\mu},n^{\mu}) = 0$. Hence, by choosing $N(x) = x$, we have made the Hamiltonian constant along flow lines as we explained before. Thus, as a result of canceling the local Weyl anomaly in a $z=1$ (1+1)-dimensional Lifshitz field theory, the \textit{cotangent} bundle of the spacetime manifold is a symplectic manifold with a Hamiltonian function $N(x) = x$ and a Hamiltonian vector field $n^{\mu}$.
	
	\section{Discussion and Outlook}\label{Discussion}
	
	\subsection{Edge Physics of the NRSCS Action}
	It is well known that the Floreannini-Jackiw (FJ) action \cite{FJ87} describes massless chiral self-dual edge bosons for the Abelian Laughlin fractional quantum Hall (FQH) state \cite{WenQFT,FradkinQFT,TongNotes}. In fact, it is the Wess-Zumino-Witten (WZW) low-energy \textit{boundary} CS action for the Laughlin state. The \textit{local} FJ action is given by
	\begin{equation} \label{FJ_Action}
	S_{FJ} = \int dt\,dx\ \partial_t\phi\partial_x\phi - v_x(\partial_x\phi)^2 \,,
	\end{equation}
	with equation of motion 
	\begin{equation}
	\partial_t\partial_x\phi - v_x\partial_x^2N= \del_t \rho (x,t) - v_x\del_x\rho(x,t) = 0 \,, \label{FJEOM}
	\end{equation} 
	where $\rho(x,t) \equiv \del_x \phi(x,t)$ is the chiral boson \textit{excitation} expressed as spatial derivative of the gauge degree of freedom $\phi(x)$. This equation has solutions of the form $\phi(x+vt)$ which describes a chiral wave propagating with constant velocity $v_x$. Replacing $\phi(x,t)$ with $N(x,t)$, $\rho(x,t)$ with $R_{xt}(x,t)$ and $v_x$ with a constant $N^x$, the FJ action becomes 
	\begin{equation} \label {}
	S_{FJ} = \int dt\,dx\ \partial_tN\partial_xN - N^x(\partial_xN)^2 \,,
	\end{equation}
	with an equation of motion
	\begin{equation}\label{chiralEOM}
	\partial_t R_{xt}(H) - N^x\partial_x R_{xt}(H)=0 \,.
	\end{equation}
	We observe that while the first term of (\ref{chiralEOM}) is the 1+1 Lifshitz Weyl anomaly, a trivial descent cocycle in the parity-odd, mixed-derivative sector of the cohomology of the Lifshitz Weyl operator, the second term $\partial_x R_{xt}(H)$ is a \textit{coboundary} term that belongs to the parity-even two-spatial derivatives sector. It is interesting to note, as pointed out in \cite{Maio2000}, that in the FJ action, it is as if the chiral boson is propagating in \textit{curved} spacetime with background metric $N^x$.
	
	Note that in deriving the boundary CS action in (\ref{FJ_Action}) from the tCS action (\ref{tCSAction}), one usually works in Galilean-boosted coordinates where the temporal component of the gauge field $a_t$ is set to zero (see equations 6.7-6.9 in \cite{TongNotes}. By doing so, one also sets the velocity of the chiral boson $N^x$ to zero and hence the chiral boson $\rho(x,t)$ is stationary, i.e. with equation of motion $\del_t \rho(x,t)=0$. Analogously, in the process of making the TTNC geometry dynamical, there is complete freedom in deciding the value of $a_t = N^x a_x +  N^r a_r$ which fixes the special conformal transformation in the $SL(2,R)$ subgroup of the Schr$\mathrm{\ddot{o}}$dinger algebra \cite{ObersDynmicTNC2015}. Choosing $N^x = N^r = 0$ directly produces the Lifshitz Weyl anomaly in (\ref{AnomnalyTorsion}). On the other hand, setting only $N^r = 0$ with a constant $N^x$ amounts to a boundary condition where $a_t = N^x a_x$ which then adds the \textit{coboundary} term $(\del_xN)^2$ to (\ref{chiralEOM}) and gives the FJ action in (\ref{FJ_Action}). We would like to further understand the relationship, if any, between the Weyl anomaly of the the $z=1$ Lifshitz theory and the FJ action in the context of the FQHE.
	
	\subsection{Anomaly Cancellation by Anomaly Inflow}
	In light of the previous discussion and deriving the 1+1 Lifshitz Weyl anomaly from a 3D non-relativistic Abelian tCS action in Section (\ref{AnomalyDerivation}) leads one naturally to wonder if the Weyl anomaly is actually somehow related to chiral edge states of a FQH theory. We discuss this possibility here. According to the classification in \cite{Fradkin15,  AbanovBoundary}, four distinct CS terms can appear in the low-energy effective action of the QH state for a microscopic theory with the following symmetries: (1) $ U(1) $ gauge transformations, (2) general covariance, and (3) local $ SO(2) $ rotations.  Written in terms of differential forms, the four CS terms are
	\begin{equation}\label{CSFQH}
	S_{CS} =  \frac{\nu}{4\pi}\int_{\mathcal{M}} A \wedge dA 
	+ 2 \bar{s} A \wedge d\omega + \overline{s^2} \omega \wedge d\omega 
	+ \frac{c}{96\pi}\int_{\mathcal{M}} \Gamma \wedge \Gamma \wedge \Gamma \,,
	\end{equation}
	where $\Gamma^{\mu}{}_{\nu} \equiv \Gamma^{\mu}{}_{\nu\rho}dx^{\rho}$. The first term is the $ U(1) $ electromagnetic Hall conductance term, while the second and third are known as the \textit{Wen-Zee} terms, and the last is the gravitational Chern-Simons (gCS). On a manifold $ \mathcal{M} $ with a boundary, the four CS terms appearing in $ S_{CS} $ defined above are no longer invariant gauge-invariant because boundary terms spoil gauge-invariance. According to \cite{AbanovBoundary}, there are then two possibilities for each CS term: (1) it represents a relevant anomaly of the low-energy effective action that cannot be canceled by adding local boundary terms, or (2) it is a \textit{trivial} anomaly which can be canceled by adding local boundary terms. The electromagnetic Hall conductance and relativistic $gCS$ terms belong to the first. The electromagnetic pure CS term can thus be made invariant as follows
	\begin{equation}
	\label{Anomalyinflow}
	\delta_{\sigma} S_{edge} = - \delta_{\sigma} S_{bulk} 
	= - k\int_{\partial\mathcal{M}}d^2x\, \sigma\,\epsilon^{\alpha\beta}\partial_{\alpha} A_{\beta} \,,
	\end{equation}
	where $\sigma$ is the gauge transformation parameter. This is an example of \textit{anomaly inflow} \cite{Harvey85} where there is an influx of charge into the boundary where they are absorbed by the anomalous gapless edge modes and as a result, the anomaly of the boundary theory gets \textit{canceled} by the total derivative term of a CS action. The Lifshitz Weyl is precisely of that nature
	\begin{equation}
	\delta_{\sigma} W_{edge} = - \delta_{\sigma} S_{tCS} 
	=  -k \int_{\del \mathcal{M}} \sqrt{h} \ N\ \sigma \  dt dx\ \left( \epsilon^{ij}\ \del_i a_j
	\right) \,.
	\nonumber
	\end{equation}
	Since the (1+1) Lifshitz Weyl anomaly, as we discussed in Section \ref{LifWeylAnomGeom} is non-trivial, in the sense that it cannot be canceled by a local boundary term, then according the classification by \cite{AbanovBoundary}, it belongs to the first class. Thus, if the microscopic theory of the Abelian QH state, in addition to having the three symmetries in (\ref{CSFQH}), is also symmetric under \textit{anisotropic local Weyl} transformations such that $W_{edge} + S_{tCS}$ is Weyl-invariant, could the boundary tCS term $ a\wedge da $ represent a new kind of \textit{torsional anomaly inflow} where torsional (or gravitational) degrees of freedom flow into the Weyl-anomalous boundary Lifshitz effective action? If so, what universal quantity, if any, does the coefficient of the tCS action represent? More importantly, is there is a physical scale-invariant FQH system? Does the NRSCS action contain the two WZ terms in (\ref{CSFQH})? We leave these questions for future work.
	
	Another related topic where anomaly cancellation by anomaly inflow is relevant is the thermal Hall effect. In \cite{Stone12}, it was shown that the thermal Hall current does not vanish in equilibrium and hence, \textit{Luttinger's} idea of coupling the system to a uniform gravitational field that such that the gravitational potential gradient exactly balances out the energy flux induced by a thermal gradient cannot be used and thus the thermal Hall conductance can not be determined by its gravitational counterpart as it was argued before in \cite{Ryu12}. In other words, it was argued in \cite{Stone12} that a \textit{uniform} gravitational field can not induce a bulk thermal current and thermal energy must therefore be carried entirely by the (1 + 1)-dimensional edge modes. The relationship the we did in to what we did in Section (\ref{AnomalyCancellation}), is to observe that as a result of canceling the Weyl anomaly and restoring Weyl invariance, the system is in \textit{equilibrium}. In fact, if we choose the lapse function $N = e^\psi = x$ such that $a_x = \del_x \psi = \frac{1}{x}$ and the Luttinger potential $\Phi(x) = \psi(x) = $\,log($x$), then
	\begin{equation}
	N(x) \del_x a_x = - \del_x \Phi(x) \,,
	\end{equation}
	\begin{equation}
	\frac{1}{T}\frac{\del T}{\del x} = - \frac{\del \Phi}{\del x}\,.
	\end{equation}
	if we identify the lapse function with inverse temperature and the torsion with the temperature gradient
	\begin{equation}
	N(x) = \beta(x) = \frac{1}{T(x)} = x, \quad a_x = T(x) =  \frac{1}{x} \,.
	\end{equation}
	More relevant to the work in this paper is the work in \cite{AbanovThermal} where a non-relativistic analogue of part of the work in \cite{Stone12} was introduced. The authors in \cite{AbanovThermal} coupled a (2+1)-dimensional non-relativistic field theory to a NC geometry with torsion.\footnote{Note that in \cite{AbanovThermal}, the torsion tensor is the curvature in the Hamiltonian $R_{xt}(H) = \del_x\psi e^{\psi}$} However, since TTNC geometry only couples to the energy density, they turned on the spatial components of the timelike vector field $n_{\mu}$ and $n^{\mu}$ to couple to the energy current.\footnote{As noted in \cite{Son13}, turning on the spatial components of the 1-form $n_{\nu}$ does not violate the Frobenius condition.} They proceeded then to construct the most general partition function with time-independent, local space and time translations and gauge symmetries. Using the Euclidean path integral to calculate the partition function, the authors in \cite{AbanovThermal} derived an expression for the thermal current. However, they did not discuss the possibility of Weyl-anomalous effective actions in the context of their work as was done in \cite{Stone12} where it was shown how the gravitational anomaly of the boundary-induced effective action can be canceled by the inflow of the spatial and temporal components of the bulk energy-momentum tensor computed from the three-dimensional gCS term (\ref{CSFQH}). Similarly, understanding the nature of this inflow requires calculating the operator conjugate to $a_{\mu}$ in the bulk which we leave for future work.
	
	
	Anomaly inflow has also been used to cancel the gravitational anomaly of a chiral field theory and obtain the Hawking radiation as was first discussed in \cite{Wilczek05} in order who further explored the relationship between Weyl anomalies and the thermal flux of the Hawking radiation as pointed out in \cite{Fulling77}. Using anomaly inflow, the authors in \cite{Wilczek05} found that the influx required to cancel the gravitational anomaly at the horizon is proportional to $T^2$ with $T=\frac{\kappa}{2pi}$ which is blackbody radiation with the Hawking temperature. This is interesting since, if we assume the field theory near the black hole horizon is the action in (\ref{SolodukhinAction}), then according to the discussion in Sections \ref{Weyl-Lorentz-Anom-Relation} and \ref{AnomalyCancellation}, canceling the Lorentz anomaly of this theory (which we recall can always be traded for a diffeomorphism anomaly by a local counterterm), is equivalent to canceling the Weyl anomaly in a $z=1$ (1+1) Lifshitz theory.

	\subsection{Boundary Conformal Charges}
	Recently, there has been some activity in constructing boundary conformal charges of type-B anomalies which are known to be Weyl-invariant densities i.e. constructed from the Weyl tensor and its covariant derivatives \cite{Solod15}. The boundary terms corresponding to the integrated anomalies are of \textit{Gibbons-Hawking-York} (GHY) type, the boundary term that must be added to the Einstein-Hilbert action to compensate for the normal variation of the metric on the boundary and thus have valid variational principle. All the Weyl anomalies of Lifshitz field theories found in \cite{Oz2014} are type B including the one we discussed in this paper. What is interesting about the 1+1 Weyl anomaly we study here, is that although it is type-B in a Lifshitz field theory, it is related to the dual of a type-A anomaly, the Ricci scalar, whose integral gives the Euler characteristic of the 2-manifold. In addition, the integral of the Weyl anomaly over the boundary gives a topological invariant (\ref{AnomalyTopo}). We would like to understand this relationship further as well as the nature of the boundary terms in (\ref{AnomalyTopo}) within the context of the discussion in \cite{Solod15,Padman14,Chakra17} especially Section 3 in \cite{Chakra17} where a comparison between the boundary term of the Einstein-Hilbert action and the standard GHY term has been shown to involve time derivatives of the lapse and shift functions. However, one can perhaps use the duality between the Ricci scalar $R$ and its dual $U$ to make a speculation as to what this topological invariant intuitively means. The foliation decomposition of $R$ in 1+1 dimensions involve terms with two temporal and two spatial derivatives. The GHY boundary term corresponding to $R$ is the trace of the extrinsic curvature (in the conformal gauge)
	\begin{equation}
	\chi = \int _{M} \,d^2x \, e\, R + \int_{\del M} \, d\tau \, \sqrt{h} \, \mathrm{(boundary \, \, term)} = \int _{\del M} \, d\tau \,\sqrt{h} \,\, \del_{\mu} \tilde{n}^{\mu} \,,
	\end{equation}
	where $\chi$ is the Euler characteristic of the 2-manifold and $\tilde{n}^{\mu}$ is a unit normal vector. Analogously, when $U$ is projected onto the foliation, it involves terms with mixed temporal and spatial derivatives while its corresponding boundary term is the divergence of a unit tangent vector $\hat{V}^{\mu}$ (\ref{AnomalyTopo}). A foliation-tangent vector we have at our disposal in 1+1 dimensions is the $x $-component of the acceleration $a_x$. Therefore, one can speculate that the boundary term corresponding to this new topological invariant may be related to $a_x$.
	
	For future work, constructing a concrete example of $z\geq1$ FPD-invariant (1+1)-dimensional Lifshitz field theory from which the Weyl anomaly discussed in this paper can be explicitly derived using heat kernel methods is important for a deeper understanding of the physics associated with this anomaly as well as the bulk theory. We only hope that the anomaly coefficient does not vanish after all.
	
	\section*{Acknowledgments}
	I would like to thank Israel Klich, Shinsei Ryu, Aron Wall, Oleg Lunin and Philip Argyres for valuable and insightful discussions. Special thanks go to Shira Chapman, Niels Obers, Jelle Hartong, and Lei Wang for reviewing large parts of the manuscript and giving constructive comments, suggestions and feedback. I would like to also thank Andrey Gromov for reviewing subsections 5.1 and 5.2 of this paper and providing very interesting ideas for future work.

\end{document}